\documentstyle[12pt,epsf]{article}  

\def\beq{\begin{equation}}
\def\eeq{\end{equation}}

\def\bea{\begin{eqnarray}}
\def\eea{\end{eqnarray}}

\begin{document}

\baselineskip=16pt
\begin{titlepage}
\begin{center}

\hfill CERN-PH-TH/2011-288
\vskip 3 cm

{\Large\bf Big Science and the Large Hadron Collider}
\vspace{1cm}

\end{center}

\begin{center}

{\bf Gian Francesco Giudice}

\vspace{.2truecm}

{\em CERN,
Theoretical Physics Division,\\
1211 Geneva 23, Switzerland}

\end{center}

\vskip 0.7in

\setlength{\baselineskip}{0.2in}

\centerline{\bf Abstract}
\vskip .1in

\noindent

%\begin{abstract}

The Large Hadron Collider (LHC), the particle accelerator operating at CERN, is probably the most complex and ambitious scientific project ever accomplished by humanity. The sheer size of the enterprise, in terms of financial and human resources, naturally raises the question whether society should support such costly basic-research programs.  I address this question here by first reviewing the process that led to the emergence of Big Science and the role of large projects in the development of science and technology. I then compare the methodologies of Small and Big Science, emphasizing their mutual linkage. Finally, after examining the cost of Big Science projects, I highlight several general aspects of their beneficial implications for society.  

%\end{abstract}
\end{titlepage}

\section{Introduction}

The Large Hadron Collider (LHC)~\cite{1},  the particle accelerator operating at the European laboratory CERN near Geneva, Switzerland, has already achieved remarkable results. On March 30, 2010, proton beams for the first time were smashed at the record high energy of 3.5 TeV (the equivalent of each proton being accelerated by a voltage of 3500 billion volts). More successes followed at an impressive rate, far beyond the most optimistic expectations. By the end of October 2010, the total production of proton crashes (the ``integrated luminosity", in technical jargon) was almost 50 inverse picobarns, the equivalent of about 5000 billion proton collisions. The conversion of the LHC to the phase in which the colliding beams were made of lead ions, instead of protons, was rapid and smooth. This operation permitted a four-week period of data accumulation that provided new information on the behavior of matter at high density. A new phase of proton collisions at high intensity began in March 2011 and on April 22, 2011, the LHC set a new record for proton-beam intensity (previously held by the Tevatron at Fermilab in Batavia, Illinois) at $4.6 \times 10^{32}$ per square centimeter per second (equivalent to about 50 million collisions per second). A few weeks later, this value was almost doubled. The LHC detectors have performed stunningly, recording with staggering precision and efficiency the mountain of data coming from the collisions. At present, the LHC has entered the phase of direct exploration of phenomena never before studied. Every indication is that new discoveries are imminent. 

	On the eve of new physics results, I will address a question that, although foreign to the immediate research goals of the LHC, necessarily concerns every large scientific project that requires enormous financial, technological, and intellectual investments -- the phenomenon of Big Science~\cite{2}.  The magnitude, complexity, and profundity of the aims of the LHC project arouse admiration and awe in the majority of people who learn about it. Nevertheless, doubts, misgivings, and even fear sometimes surface both outside and inside the scientific community regarding anything associated with Big Science. The question therefore is: Should society support large research projects in basic science? 

\section{The Emergence of Big Science}
 
The Manhattan Project is often considered to be the beginning of Big Science, establishing a new and tighter relationship between science and society, and creating a new methodology in scientific investigation. Aside from any moral considerations, it is undeniable that the Manhattan Project defined the {\it modus operandi} of Big Science, a trademark with the following definite characteristics.  A large number of scientists is involved in a project whose target is well defined, although it requires crossing the limits of known science and technology. Large funds are made available to the project, but the goal has to be reached within an established period of time.  Scientists must adapt to work in interdisciplinary groups which, in the case of the Manhattan Project, mixed theoretical and experimental physicists with engineers and mathematicians.  Finally, the project is placed under the direct control of administrative bodies external to its academic environment. 
 
	In reality the Manhattan Project was only an episode that accelerated an already inevitable evolutionary process. Well before the Second World War, rapid scientific and technological progress had propelled science beyond its academic borders. On the one hand, science was having an increasingly crucial impact on society; on the other hand, science was requiring financial resources that could be found only outside of the limited world of universities and research institutions.  The construction of increasingly more advanced and costly instruments was becoming decisive for progress in many scientific fields. 
 
	One example is the field of stellar astronomy, which required more and more powerful optical telescopes. The race for cutting-edge instrumentation led to the completion of the legendary 100-inch Hooker telescope at Mount Wilson Observatory in 1917, with which Edwin Hubble discovered that the Andromeda nebula is much more distant from us than the boundary of the Milky Way, thus proving that our galaxy is only one of a multitude of galaxies that dot the night sky. This discovery changed forever our vision of the universe. With this same instrument Hubble also made his famous observations on galaxy recession, thereby proving that our universe is expanding. In the absence of the planning that led to the construction of the 100-inch Hooker telescope, these revolutionary discoveries would not have been possible.  Although powerful and expensive optical telescopes were used in astronomy, the observations involved only small groups of scientists and hence did not have all of the typical characteristics of Big Science. That changed later with the advent of radioastronomy. 
 
	A second example was the race for ever lower temperatures, which from its inception required more and more advanced and complex instruments. The main protagonists in this race were James Dewar, who carried out his experiments at the Royal Institution in London, and Heike Kamerlingh Onnes of the University of Leiden. Kamerlingh Onnes, a great experimentalist with a decidedly pragmatic inclination, organized his laboratory almost like a factory (jokingly called ``The Brewery") and obtained some of his funding from the refrigeration industry. His organizational skills were instrumental to his success. On July 10, 1908, he liquefied helium, the last element that was still known only in the gaseous state.  To produce a small 6-centiliter vial of liquid helium he had to reach the record temperature of $-$269 degrees Celsius, about 4 degrees above absolute zero. This result paved the way to the later discovery of superfluidity. In the meantime, in 1911, Kamerlingh Onnes used liquid helium to cool mercury and discovered the astonishing phenomenon of superconductivity, according to which some materials completely lose their electrical resistivity below a well-defined critical temperature. 
 
         I note in passing that the physical phenomena associated with superconductivity and superfluidity are crucial for the functioning of the LHC. Inside its underground tunnel 1200 tons of superconducting cables transport the exceptionally intense electric currents (up to 12,800 amperes) that generate the magnetic fields used to guide the proton beams. Dipole magnets are aligned along the 27-kilometer tunnel, and superfluid helium is employed to maintain them at the temperature of $-$271 degrees Celsius (1.9 degrees above absolute zero). Without knowledge of superconductivity and superfluidity the LHC would not have been imaginable. 
 
	A third example was the quest to probe the internal structure of the atom, which required ever more expensive equipment and tools.  In particular, the rising cost of radium after its discovery by Marie and Pierre Curie in 1898 greatly restricted the number of universities and laboratories that could afford to support research on atomic and nuclear structure. Radium was commonly used as a source of alpha particles to probe the atom, but it reached the cost of \$160,000 per gram, becoming by far the most precious substance in the world. 
 
	In the wake of continually rising costs, the importance of managerial capabilities of scientists also grew. In the early twentieth century, some research funds were provided by industrialists, philanthropists, and other benefactors, especially in Anglo-Saxon countries.  Later, when it became necessary to also turn to the public sector, scientists encountered the need to communicate their research and its results to the general public, which often was fascinated with them.   Albert Einstein was an undisputed cultural icon, but even the far less dazzling Paul Dirac was able to attract crowds. When he gave a public lecture on a cricket field in Baroda, India, thousands of people came, making it necessary to use a cinema screen for those unable to get inside the stadium~\cite{3}.  The much more difficult problem was to find support among politicians and administrators. In the field of atomic and nuclear research, Ernest Rutherford in Great Britain and Ernest Orlando Lawrence in the United States excelled in being able to secure both private and public research funds.
  
	Yet another link, a military link, was established between science and government during the First World War. Chemistry was at the forefront in designing and producing chemical weapons. In August 1914, the French army employed tear gas for the first time, while at the battle of Ypres in April 1915 the German army used poisonous gases made of chlorine, phosgene, and yperite (named after the Belgian city of Ypres, but commonly known as mustard gas).  The Allied forces initially reacted by condemning the German action, but then began to develop their own research programs in chemical weapons, which were first used toward the end of 1915. It is estimated that chemical warfare caused serious, often permanent injuries to more than a million soldiers on both sides of the conflict, killing 90,000 of them (out of which 56,000 were Russian). Physics participated in the war with wireless communication, a new means for coordinating military field actions, and with instruments for detecting submarines by means of acoustic techniques, the precursors of sonar. Even pure mathematics was not immune, since in the hands of the military cryptography became yet another weapon.  Moreover, many scientists and engineers participated in wartime committees, finding themselves sitting around the same table with military officers, politicians, and industrialists. This liaison paved the way for a new role of science in society.

\section{A New Science-Society Environment}

During the Second World War, science was involved in two major projects, the development of radar, in which the United States invested 3 billion dollars, and the Manhattan Project~\cite{4}, which cost about 2 billion dollars.   The Manhattan Project, a dreadful scientific challenge fueled by the fear that the Germans would beat the Allies in building an atomic bomb, represented a decisive step in the evolution of Big Science because it introduced a special methodology, rather unusual for the traditional standards of scientific research at that time.

	At the end of hostilities, the United States awakened from the nightmare of war with an unshakable faith in science. Physicists, seen as the essential contributors to military supremacy, enjoyed special consideration, and particle physics, seen as the heir of the science that led to the Manhattan Project, became one of the main beneficiaries of public financial support. The Cold War helped to cement this privileged status, but many particle physicists regarded the sympathy of military circles as an ethically uncomfortable legacy, though a profitable one. Physicists who during the war had become ``better scientists if impurer men"~\cite{5},  now sought a form of redemption by working on peaceful applications of nuclear energy, or in exploring the secrets of nature at the subnuclear level. 
	
	This postwar favorable environment affected science as a whole. The United States witnessed an explosion of publicly-funded scientific projects. Many U.S. economists, influenced also by the theories of the Austrian-American economist and political scientist Joseph Schumpeter, saw scientific research and technological innovation as keys to constant economic growth, producing jobs and prosperity.  As a consequence, it also implied a solution to the social problems of the poorer classes and a preemptive cure to possible political instability. Basic science and scientific research were essential links in this logical chain. 
	
	Vannevar Bush vigorously asserted these links in his influential report, {\it Science: The Endless Frontier}, which he submitted to President Harry S. Truman on July 5, 1945:  ``The simplest and most effective way in which the Government can strengthen industrial research is to support basic research and to develop scientific talent"~\cite{6}.   Bush thus identified basic research as the decisive factor in promoting progress, claiming that technology is an inescapable consequence of leading-edge science. The goal of government thus is to support and nurture the most advanced research institutions without paying much attention to aspects of technological innovation. According to Bush, ``basic research is the pacemaker of technological progress"~\cite{7}.  In the report he also proposed the creation of what later became, in 1950, the National Science Foundation. 
	
	The economist and Assistant to the President John R. Steelman asserted an analogous point of view in his report, as Chairman of The President's Scientific Research Board, of August 27, 1947: ``Only through research and more research [in the basic sciences] can we provide the basis for an expanding economy, and continued high levels of employment"~\cite{8}.   President Truman responded on September 13, 1948, announcing the main points of his program for scientific development: ``First, we should double our total public and private allocations of funds to the sciences... 
	Second, greater emphasis should be placed on basic research and on medical research. 
	Third, a National Science Foundation should be established. 
	Fourth, more aid should be granted to the universities, both for student scholarships and for research facilities. 
	Fifth, the work of the research agencies of the Federal Government should be better financed and coordinated"~\cite{9}.
	  
During this period of unprecedented expansion of American scientific research, an event occurred that galvanized public attention and undermined the government's conviction of possessing complete technological hegemony.  On April 12, 1961, Yuri Alekseyevich Gagarin became the first man to journey into space.  The United States reacted immediately. On May 25, 1961, President John F. Kennedy addressed the U.S. Congress and nation with the famous words: ``I believe that this nation should commit itself to achieving the goal, before this decade is out, of landing a man on the Moon and returning him safely to Earth".  Public opinion was firmly in his favor, and Congress showed no hesitation, almost unanimously approving this colossal project, estimated to cost between 20 and 40 billion dollars. Without entering into the debate about their scientific value, the Apollo missions exhibited the characteristic {\it modus operandi} of Big Science, though in a context very different from that of the Manhattan Project. Further, the alleged ``missile gap" (the perceived technological lag of the United States behind the Soviet Union) had to be closed as rapidly as possible.  Remedies were sought not only in the space race but also in basic research and in education, for example by strengthening school curricula in science and mathematics.
 
	In this euphoric atmosphere, some expressions of doubt about large publicly-funded scientific projects began to emerge not only in society, but also within the scientific community.  The most authoritative voices included physicists Merle A. Tuve, Alvin M. Weinberg, and Philip W. Anderson, and astrophysicist Fred Hoyle.  In 1961 Weinberg\footnote{Alvin M. Weinberg should not be confused with the theoretical physicist Steven Weinberg, who told the following story: ``In 1966 when I was first visiting Harvard I found myself at lunch at the faculty club with the late John Van Vleck... Van Vleck asked me if I was related to `the'  Weinberg. I was a bit put out, but I understood what he meant; I was at that time a rather junior theorist, and Alvin was director of the Oak Ridge National Laboratory. I dipped into my reserves of effrontery, and replied that I was `the' Weinberg. I do not think that Van Vleck was impressed"~\cite{9bis}.}, who since 1955 was Director of Oak Ridge National Laboratory,  which had supplied the enriched uranium for the Manhattan Project, published an influential essay on the impact of large scientific projects in which he coined the term ``Big Science"~\cite{10}.  He wondered whether Big Science was ruining science, identifying some issues that are still worth discussing today. 
``In the first place, since Big Science needs great public support it thrives on publicity. The inevitable result is the injection of a journalistic flavor into Big Science which is fundamentally in conflict with the scientific method... The spectacular rather than the perceptive becomes the scientific standard"~\cite{11}. 
Weinberg was referring to the space program; today his words bring to mind certain unfortunate kinds of information about the LHC promoted at times by CERN. 

	The enormous size of Big Science projects requires constant oversight by administrative bodies, which Weinberg saw as an abandonment of true scientific motives: ``Unfortunately, science dominated by administrators is science understood by administrators, and such science quickly becomes attenuated if not meaningless"~\cite{12}. The true risk is excessive bureaucratization of large scientific projects. Public authorities, which have the fair duty of monitoring the expenses incurred by large projects, can impose decisions based on purely financial considerations, neglecting their scientific and technical aspects.  Administrators are accustomed to operate quite differently than scientists, and can even inadvertently destroy the special vitality that thrives in a research environment. More than three decades after Weinberg wrote, Wolfgang K.H. (ÒPiefÓ) Panofsky, the exuberant physicist who since 1961 served for twenty-three years as Director and subsequently as Director Emeritus of the Stanford Linear Accelerator Center, identified bureaucratization as the main cause that led to the cancelation of the Superconducting Super Collider (SSC) in October 1993. 
``The sheer size of the undertaking, the micromanagement by DOE [the Department of Energy], and the intensity and frequency of external oversight all led to a bureaucratic internal culture at the laboratory. In the name of cost control, technically needed changes and design trade-offs were discouraged. Decisions on technical alternatives were distorted by ``political acceptability" and were at times made late or not at all... Key scientific and technical people were generally placed low in the decision chain"~\cite{13}.  

In his aforementioned article, Weinberg performed a calculation that today makes us smile (or perhaps frown). He extrapolated the rate of growth of the cost of scientific research since the end of the war to 1961 and concluded that in another twenty years science would ruin the United States financially. This danger has clearly been averted, but his concern gives us a measure of the exceptional U.S. investments in research during the postwar period. Weinberg's worry was also a harbinger of the disillusionment toward science that accompanied the social transformations and political and ideological movements of the 1960s and 1970s. The public became aware that technology brings not only progress, but can also lead to social injustice and environmental damage.  The philosopher Herbert Marcuse, who much influenced the generation of the 1968 protests, maintained that science, by its very nature, induces an inhumane way of thought and that technology is an engine of oppression.  We encounter here a typical limitation, namely, the inability to distinguish clearly between science and technology, and to identify and associate both with war. At the same time, the Vietnam War, besides fueling popular discontent, revealed the limits of advanced military technology. The U.S. army, despite its sophisticated weaponry, was kept in check by the poorly-equipped but resolute North Vietnamese army. Moreover, the expenditure of large public funds began to weigh upon the internal budgets of Western countries. The capability and will to sustain large scientific projects began to dwindle. 

	The fall of the Berlin Wall in November 1989, and the consequent dissolution of the Soviet Union in 1991, dissipated the specter of the Cold War, and with it the motivation of national prestige that influenced political support for large scientific projects. In October 1993 the U.S. Congress canceled the SSC, the accelerator that would have collided protons with an energy almost three times higher than that of the LHC and that had been approved more than six years earlier. Many are the reasons that led to its unfortunate cancellation, after its construction had already cost almost 2 billion dollars~\cite{14},  but I will mention only one issue that, although not necessarily the most important one, is particularly germane to my discussion. The SSC project was approved during the Reagan Administration, in an era of revived public spending, but also when concerns about national defense took precedence over ones about science.  The Strategic Defense Initiative (``Star Wars"), then estimated to cost around 60 billion dollars, and the Freedom Space Station were approved by Congress in this period. The SSC was canceled under the Clinton Administration after the end of the Cold War and, more importantly, at a time when Congress was determined to reduce the increasing U.S. public deficit. It is worth noting that just two days before voting to cancel the SSC, the U.S. House of Representatives had expressed support, albeit by a majority of but a single vote, for the continuation of the International Space Station (a synthesis of the Freedom Space Station and similar projects initiated by Russian, European, and Japanese space agencies). At the time, the International Space Station was estimated to cost more than three times that of the LHC, its cost was continually rising, and the scientific motivations for its construction were rather weak. The international element and the prior agreements with foreign countries certainly worked in favor of the International Space Station. 
	
	The cancellation of the SSC was a traumatic event for the particle-physics community around the world. It marked the end of an era, but not the end of large basic-science projects. It represented an important step in the evolution of Big Science, because it highlighted the need for new characteristics in large scientific projects. A broad international collaboration, and a vision beyond the interests of any single country, proved to be essential elements for their success. The LHC, built by a consortium of the European member states of CERN with substantial contributions from almost all of the main countries in the world, has achieved this vision superbly. 

\section{Big Science and Small Science}
 
The charm of science is usually associated with the image of the brilliant idea, born in the silence of a sleepless night and elaborated with mathematical calculations in a simple notebook: the creation of an individual that revolutionizes the foundation of our understanding of nature.  Or we imagine a scientist who, in the solitude of a laboratory, designs and performs an extraordinary experiment, discovering a new and completely unexpected phenomenon. At first sight, Big Science seems to be the antithesis of these images. 

	This contrast does not necessarily correspond to reality. As we have seen, the development of a leading-edge scientific field leads inexorably to the need for large undertakings and ambitious projects. Even fields that have been regarded traditionally as Small Science (such as molecular biology or climate science) today require programs with typical Big Science features (such as the Human Genome Project or supercomputing for studying climate changes). These large-scale projects are not necessarily the antithesis of the poetic and traditional view of science, but rather are its natural completion and enrichment. The two methods of investigation are not opposed, because both share the same scientific ethic and ultimate goal. Both methods are necessary for science to advance beyond its current limit of knowledge. It is like comparing a painting by a Renaissance master with the epic construction of a Gothic cathedral.  The advancement of art requires both. 
	
	Whether we like it or not, Big Science is an irreplaceable instrument of modern science. Wherever science makes progress, sooner or later the need for large and expensive instruments, for goal-driven organized undertakings, for tight collaboration of scientists educated and trained in different disciplines, will arise. The duty of scientists and of their funding agencies is to employ wisely the special instrument of Big Science for projects of indisputable scientific excellence, free from motivations of national prestige or propaganda, and devoid of any military motivation. 
	
	A common criticism of Big Science is the claim that it transforms research from a method of scientific investigation into an industrial process that stultifies creativity. In reality, Big Science is only a technical necessity and not a dismantlement of traditional scientific goals, values, and motivations. The methods of investigation have changed, but not the principles and passions that drive scientists. Enrico Fermi provides an excellent example. The great Italian physicist experienced all of the various ways of doing science: the pensive and individualistic style of theoretical physics (with his statistics of half-integer spin particles and his theory of beta decay), the spontaneity and enthusiasm of doing Small Science (with his slow-neutron experiments that he and his {\it Via Panisperna} Boys carried out in the goldÞsh pond of the physics department garden), and the goal-driven and organized structure of Big Science (with the Chicago Pile and Manhattan Project). A Big Science project also thrives on individual creativity, for which the LHC provides ample evidence. 
	
	Another criticism of Big Science originates from the conflict between two epistemologically different positions which, depending upon the context, have been called ``intensive research" and ``extensive research"~\cite{15},  or ``reductionism" and ``constructionism"~\cite{16},  or ``fundamentalist" and ``generalist" physics~\cite{17}.   It is an empirical fact -- not a philosophical assertion -- that nature exhibits an ordering, at least up to the distances explored until now. Simpler elements emerge at smaller distances. Furthermore, the physical laws that govern the simplest elements reveal fundamental and universal properties. These physical laws allow us not only to understand the particle world, but also to describe the large-scale structure of the universe and to reconstruct its history since its very first instants. Reductionism aims at discovering these laws and is driven by human curiosity to comprehend the ultimate workings of nature. 
	
	Knowledge of the fundamental physical laws, however, is often not sufficient to describe, from a practical and quantitative point of view, the complexity of many natural phenomena. In other words, knowledge of an equation does not imply the capability to derive a solution that is suitable for describing a particular phenomenon. The mathematical description of the emergent properties of a complex system requires physical laws that are completely different from those of the fundamental theory. Here enters constructionism, which aims at discovering these emergent laws. 
	
	Both programs, the reductionist and the constructionist, have their scientific validity and intellectual interest. The mere existence of these two different approaches demonstrates the richness and variety of the scientific panorama. It would be dangerous to claim that all scientific research should follow a single path. 
	
	The distinction between research in reductionist fields (high-energy physics, cosmology) and constructionist fields (solid-state physics, astronomy, molecular biology) no longer corresponds to the distinction between Big Science and Small Science since both sectors have developed large projects. Moreover, the distinction between reductionism and constructionism seems to be linked to the image of a certain field at a particular historical moment, without corresponding to a real difference in the basic motivations of the scientists active in research. For instance, in the past, nuclear physics was considered to be a reductionist science, but today it no longer is, and in astronomy the reductionist activity of observational cosmology coexists with a constructionist soul.  This seems to indicate that this distinction is more significant for historians of science than for scientists themselves. 
	
	This problem of semantics acquires a less academic flavor when different research sectors compete for public funding. A commonly expressed fear is that Big Science projects could absorb all available resources, suffocating the activities of smaller and less organized sectors. In principle, this worry is valid because diversification of research is vital for scientific development. In practice, however, public funding for science is never a simple zero-sum game.  Decision-making mechanisms are more complicated, and the approval of large projects is not necessarily in conflict with a robust program of scientific diversification. As a matter of fact, in the past, large and small projects have always risen and fallen together.  For example, there is no evidence that after the SSC was cancelled the scientific fields whose representatives openly opposed its continuation enjoyed any financial benefit. 
	
	If a distinction must be made, it is better to make it between projects and directions of research that drive real advancement in knowledge and those that lead to blind alleys or propose repetitive experiments of little scientific value. Science needs different investigational methodologies to create the opportunities favorable for making progress. Freeman Dyson concluded one of his anti-SSC sermons by asserting that 
``there is no illusion more dangerous than the belief that the progress of science is predictable. If you look for nature's secrets in only one direction, you are likely to miss the most important secrets, those which you did not have enough imagination to predict"~\cite{18}. 
Dyson's perceptive words, however, do not necessarily undermine the reasons for big scientific enterprises. Even such an adamant Small Science advocate and Big Science critic as Dyson concurs that the extraordinary progress in astronomy and particle physics during the last sixty years was made possible only by a wise admixture of large and small projects. There were triumphs and unexpected discoveries, as well as failures and errors, in both Big Science and Small Science, but final success could never have been achieved without the existence of both large and small projects. The stability of an ecosystem needs animals of different sizes, but the size of an animal species does not establish its aptness for survival, which instead is determined by the interrelationships of bigger and smaller creatures. So it happens in science: there can be no long-term growth in a system where large projects absorb the totality of resources, nor where prejudicial objections are pitted against large projects.

\section{The Cost of Big Science Projects}
 
Are large basic-science projects too expensive or even useless? Representative Sherry Boehlert (Republican, New York), a tenacious opponent of the SSC, addressed the House in 1992 during a discussion of its motivations, asserting that ``the SSC will not cure cancer, will not provide a solution to the problem of male pattern baldness, and will not guarantee a World Series victory for the Chicago Cubs"~\cite{19}.   I cannot disagree. But to address the question of whether or not society should embark upon large basic-science projects it may be more appropriate to consider other arguments. 

	I first analyze one aspect of the issue of costs that is related to the management of large scientific projects by nonscientist administrators, a point that Alvin Weinberg raised already in his 1961 article. A budget inflexibility that fails to allow for studying alternatives and for dealing with contingencies is dangerous policy. Equally dangerous is rigidity in commitment to an initial project design in the face of new technological or scientific developments. Such policies, for the sake of cost control, can turn into far greater financial losses for a project or even into its scientific failure. According to Dyson, the setback of the Shuttle Program was caused largely by premature choice that was imposed upon NASA engineers. 
``Premature choice means betting all your money on one horse before you have found out whether she is lame. Politicians and administrators responsible for large projects are often obsessed with avoiding waste. To avoid waste they find it reasonable to choose one design as soon as possible and shut down the support of alternatives. So it was with the shuttle... The evolution of science and technology is a Darwinian process of the survival of the fittest. In science and technology, as in biological evolution, waste is the secret of efficiency. Without waste you cannot find out which horse is the fittest. This is a hard lesson for politicians and administrators to learn"~\cite{20}.

\begin{table}[h]
\centering
\begin{tabular}{|c||c|c|}
\hline
& Original & Estimated \\
Project& cost & cost in 2011 \\
& $\times 10^9$ dollars & $\times 10^9$ dollars \\
\hline
\hline
{\bf Manhattan Project}~\cite{22}  & &\\
Estimated cost at approval (1942): 3 years 1942-1944 & 0.148 &\\
Total cost: 5 years 1942-1946	& 2.2 & 27\\
\hline
{\bf Apollo Program}~\cite{23}  & &\\
Estimated cost (1966): 13 years  & 22.7 &\\
Total cost: 14 years 1960-1973	& 19.4 & 120\\
\hline
{\bf Hubble Space Telescope (HST)}~\cite{24}  & &\\
Initial estimated cost  & 0.5 &\\
Construction cost  & 1.5 &\\
Total estimated cost: 25 years 1990-2014	& 6.0 & 8\\
\hline
{\bf Superconducting Super Collider (SSC)}~\cite{25}   & &\\
Estimated cost at approval (1987) & 4.4 &\\
Estimated cost at cancellation (1993) & 11.8 &18\\
\hline
{\bf International Space Station (ISS)}~\cite{26}    &  &\\
Initial estimated cost  & 17.4 &\\
Estimated cost for development, assembly and operation (1998) & 96 &120\\
\hline
{\bf Human Genome Project (HGP)}~\cite{27}   & &\\
Scientific program in genomics total cost: 14 years 1990-2003 & 3 & 4\\
\hline
{\bf International Thermonuclear Experimental} & & \\
{\bf Reactor (ITER)}~\cite{28}   &  &\\
Estimated construction cost (2010): 10 years 2008-2017 & 17.9 & 18\\
\hline
{\bf Large Hadron Collider (LHC)}   &  &\\
Materials for construction of accelerator and detectors & 5.4 & 6\\
\hline
\end{tabular}
\caption{{Original cost estimates of some Big Science projects in billions of dollars and their equivalents in billions of 2011 dollars~\cite{21}.  I used the following conversion factors: 1 euro = 1.4 dollars = 1.5 Swiss francs. I chose an average value of the Swiss franc at the time of construction of the LHC rather than today's exchange rate.}} 
\end{table}

To form an opinion on the choices society must face regarding big scientific enterprises, it is useful to review their costs.  I summarize the costs of some large projects in table 1. The data should be interpreted with great caution, because the way in which costs are estimated varies enormously from project to project and, in some cases, a single program shares so many different aspects of research and technology that it is virtually impossible to reliably quantify its expenses.  Note that the expenses for the Manhattan Project (which amounted to 0.6\% of the U.S. military expenses during the Second World War) were mostly for the plants that separated the uranium isotopes and for the production of plutonium at Oak Ridge and Hanford. The Los Alamos laboratory, where most of the physicists worked, cost only 4\% of the total expenditure for the Manhattan Project. 

	I summarize the LHC costs, according to the CERN budget, in table 2. I should note that only people directly employed by the LHC are counted under the heading of ``Personnel", but in practice a large fraction of CERN personnel works for the LHC. Moreover, the data do not include the costs of operation and the contributions to the construction and functioning of the particle detectors made by universities and laboratories outside CERN. For example, the material costs of the largest detector (ATLAS) were 540 million Swiss francs, and CERN contributed to the costs of the various detectors an amount that varied between 14\% and 20\% of the total. 
	
	\begin{table}[h]
\centering
\begin{tabular}{|c||c|c|c|}
\hline
 & Personnel & Materials & Total \\
 & $\times 10^9$ Swiss& $\times 10^9$ Swiss&$\times 10^9$ Swiss\\
 & francs & francs & francs\\
 \hline
 \hline
 LHC machine and experimental areas & 1.224 & 3.756 & 4.980\\
 (incl. R\&D, injectors, tests and pre-operation) &  &  &\\
 \hline
 CERN contribution to detectors & 0.869 & 0.493 & 1.362\\
(incl. R\&D tests and pre-operation) &  &  &\\
\hline
CERN contribution to LHC computing & 0.085 & 0.083 & 0.168 \\
\hline
Total CERN costs & 2.178 & 4.332 & 6.510\\
\hline
\end{tabular}
\caption{{Costs of the LHC in billions of Swiss francs according to the CERN budget~\cite{29}.}}
\end{table}	

	Just for the sake of comparison, the LHC costs roughly as much as large public civil-engineering projects. For example, the 8-kilometer {\O}resund bridge, which was completed in 2000 to connect Denmark to Sweden, cost about 4 billion euro, and the 40-kilometer bridge over the Strait of Messina, which may one day connect Sicily to the rest of Italy, is today estimated to be around 6 billion euro, but that presumably will increase. The estimated cost of the 2012 Olympic Games in London have already passed 10 billion euro. 
	
	It is difficult to set an accurate and meaningful price on the value of knowledge, on the cultural impact of scientific discoveries, on the human desire to understand the organizing principles of nature and to decipher the universe. It is easier, however, to identify a reciprocal causal link between progress in knowledge and economic, social, and industrial development. Each strengthens the other in a symbiotic relationship, as occurred in Great Britain at the end of the nineteenth century, in Germany at the beginning of the twentieth century, and in the United States after the Second World War. Advancements in basic science rarely have immediate effects on technology; their value for society appreciates with time. Today's technological sectors were subjects of past basic research. 
	
	The only legitimate yardstick for measuring the importance of a basic-science project is its impact on science itself. Economic and technological relevance does not always lead to best-science choices and thus does not always translate into a better investment for society. Nevertheless, the enormous costs of large scientific projects justify accurate analyses of possible economic and technological spinoffs on the part of the funding agencies. These evaluations depend on the specific cases in question, but Big Science projects present some common features that often translate into special opportunities. Quite independently of the scientific goals of a project, which are its true and only {\it raison d'\^etre}, I can highlight some general aspects of Big Science that can have beneficial effects on society. Needless to say, one perhaps could make an equally long list of arguments in support of the social advantages of the Small Science approach. 
	
	1. The large concentration not only of financial, but especially also of intellectual resources in the same research center creates an environment that hardly can be achieved in academic institutions. This produces very fertile ground that is naturally open to innovation, well beyond the planned objectives of the project. For example, it was not by chance that the worldwide web was born at CERN, even when its creation was not one of the direct goals of the laboratory. 
	
	2. From the point of view of applications, the fruits of basic research are usually slow to ripen. This lag between a scientific discovery and its technological spinoffs is filled naturally by the methodology of Big Science because, while applications of the ultimate goals of large projects are nearly impossible to foresee, the real technological significance of the project lies in the research developed to accomplish these goals. The LHC provides an excellent example. No one can say with certainty today if and how the discovery of the Higgs boson, or of any other exotic particle, will plant the seed for some practical application. But the research that led to the construction of the LHC already has translated into many useful spinoffs:  accelerator development has produced hadrotherapy for treating cancer and synchrotron light with its many uses as an ``X-ray microscope"; particle-detector development has produced various medical-diagnostic techniques and real-time analyses; information development has produced the worldwide web and grid computing. 
	
	3. The need for advanced technologies and the consequent close relationship with private companies offer benefits to the industrial sector that go beyond the simple profits assessed in terms of their contracts.  Scientists' requests for cutting-edge prototypes pushes industries to develop new manufacturing techniques whose development would be too risky in merely a market environment. 
	
	4. Basic research, because of its universal character and freedom from economic or military interests, is particularly well suited for international collaboration, and large projects are the best vehicles for it. These projects offer countries an opportunity for participation in great scientific challenges that they themselves lack the resources to pursue. Moreover, large scientific ventures can strengthen peaceful international ties, and even open up collaborations between hostile nations, thereby creating opportunities for political rapprochement. In the climate of the Cold War, Alvin Weinberg, despite his aversion to Big Science, understood its special role here:  ``If high-energy physics could be made a vehicle for international cooperation ... between East and West ... the expense of high-energy physics would become a virtue"~\cite{30}.  A laudable present-day example comes from SESAME (Synchrotron light for Experimental Science and Applications in the Middle East), a research project based in Jordan and operated by a scientific collaboration involving Israel, Iran, and other Middle East countries as well as the Palestinian National Authority. 
	
	5. Large scientific projects provide a unique opportunity for educating and training students and young researchers. For example, young people play an instrumental role at the LHC.  About half of the physicists involved in the ATLAS experiment are younger than thirty-five (and almost a third are younger than thirty). These young scientists and engineers learn how to tackle complicated problems, to master advanced technologies, and to work in interdisciplinary teams. Not all of them will remain in scientific research, but they will carry their unique skills and experience into other sectors of society. Investments in large scientific projects constitute investments in future generations of capable and competent members of society. 
	
	6. Large projects are often irreplaceable tools for the advancement of basic science. Relinquishing these tools means relinquishing precious knowledge that has a value that transcends the boundaries of science, affecting all society.  Knowledge has an intrinsic value that is linked to our awareness of the meaning of nature, and of the role we play in our physical universe. This awareness influences our way of thinking and acting as individual human beings and as a community, thereby contributing to the intellectual growth of society. In this sense, the value of basic science is not unlike that of art.  Large scientific projects, like inspiring works of art, capture the public imagination and transfer knowledge in particularly effective ways throughout all layers of society. 
	
	Every civilization, every historical epoch, leaves a legacy to future generations. I believe that the legacy of our society will be found in the revolutionary scientific discoveries that were and are being made, and in the swift technological progress that was and is being achieved through them, which have dramatically changed not only how we live, but especially the way we think and comprehend the universe.  Large scientific projects have played a catalyzing role in these profound changes, and the LHC exhibits all of the characteristics that promise it too will be remembered as such.  It should not come as a surprise that there is growing excitement, not only among physicists, but also among the general public for the upcoming LHC results, which will probe a world of matter that is apparently so foreign to our common experience, but one that hides the essence of the physical laws that govern the universe. 

\bigskip

{\bf Acknowledgments}
 
This article is based on a colloquium I gave at the Scuola Normale Superiore in Pisa, Italy, on May 5, 2010. Its content was largely stimulated by a question that Giovanni Bignami asked during a public presentation on the physics of the LHC that I gave at the Istituto Veneto di Scienze, Lettere ed Arti in Venice.  I also thank Guido Altarelli, Riccardo Barbieri, Michelangelo Mangano, Marco Martorelli, Markus Nordberg, Emma Sanders, Anders Unnervik, Gabriele Veneziano, and James Wells for useful comments and discussions. Finally, I express my deep gratitude to Roger H. Stuewer for his thoughtful and careful editorial work that has greatly improved my article.

\end{document}